\documentclass[12pt]{iopart}

\usepackage{graphicx}
\usepackage{epsfig}

\begin{document}

\title[Microscopic calculations of isospin mixing and isospin-symmetry-breaking corrections]
{Microscopic calculations of isospin mixing in N$\approx$Z
nuclei and isospin-symmetry-breaking corrections to the superallowed
$\beta -$decay}

\author{M Rafalski$^{1,2}$ and W Satu{\l}a$^1$}

\address{$^1$Faculty of Physics, Institute of Theoretical Physics,
University of Warsaw, \\ ul. Ho\.za 69, PL-00-681 Warsaw, Poland}
\address{$^2$Department of Physics \&
  Astronomy, University of Tennessee, Knoxville,\\ Tennessee 37996, USA}

\begin{abstract}
Recently, we have applied for the first time the angular momentum and isospin projected nuclear density functional theory
to calculate the isospin-symmetry breaking (ISB) corrections to the superallowed $\beta$-decay.
With the calculated set of the ISB corrections we found
$|V_{\rm ud}|=0.97447(23)$ for the leading element of the Cabibbo-Kobayashi-Maskawa matrix. This is
in nice agreement with both the recent result of Towner and Hardy [Phys. Rev. {\bf C77}, 025501
(2008)] and the central value deduced from the neutron decay.
In this work we extend our calculations of the ISB corrections covering all superallowed transitions $A,I^\pi=0^+,T=1,T_z \rightarrow
A,I^\pi=0^+,T=1,T_z+1$ with $T_z =-1,0$ and $A$ ranging from 10 to 74.
\end{abstract}

\pacs{
21.10.Hw, 
21.60.Jz, 
21.30.Fe, 
23.40.Hc 
}
\submitto{\PS}

\maketitle

\section{Introduction}

The isospin symmetry~\cite{[Hei32],[Wig37]} in a nuclear medium is only weakly broken
reflecting the relative weakness of isospin-breaking as compared to the isospin-conserving part of the nucleon-nucleon interaction.
Hence, the related isotopic spin quantum number, $T$, albeit approximate, remains very useful for labeling
nuclear states and understanding selection rules on different types of nuclear reactions (see Ref.~\cite{[Wil69]} and
refs. therein). In particular, for the Fermi (vector) and  Gamow-Teller (axial) $\beta$-decay one has $\Delta T =0$ and $\Delta T =0, \pm 1$, respectively, with the exception of $I^\pi=0^+ \rightarrow I^\pi=0^+$ transitions which are strictly
forbidden for the Gamow-Teller process~\cite{[Wig39]}.

Among pure Fermi transitions of particular importance are superallowed transitions
between the isobaric analogue states $I^\pi=0^+, T=1, T_z \rightarrow I^\pi=0^+, T=1, T_z+1 $
in $N\approx Z$ nuclei. These data are used for precision tests
of the conserved vector current (CVC) hypothesis  i.e.
the independence of the vector current on the nuclear medium.
With the CVC hypothesis being verified, they serve as
the most precise source of the $V_{\rm ud}$ element of the Cabibbo-Kobayashi-Maskawa
(CKM) matrix which is a key ingredient in investigating its
unitarity.

The CVC hypothesis is verified by investigating nucleus-independence of
the ${\cal F}t$-values:
\begin{equation}\label{ftnew}
   {\cal F}t \equiv ft(1+\delta_{\rm R}^\prime)(1+\delta_{\rm NS} -\delta_{\rm C})
  = \frac{K}{2 G_{\rm V}^2 (1 + \Delta^{\rm V}_{\rm R})} = {\rm const},
\end{equation}
where $K/(\hbar c)^6 = 2\pi^3 \hbar \ln 2 /(m_e c^2)^5 =
8120.2787(11)\times 10^{-10}$\,GeV$^{-4}$s  is
a universal constant and $G_{\rm V}$ stands for the vector coupling constant for the
semileptonic weak interaction.
The ${\cal F}t$-values include empirical reduced life-times $ft$
corrected, theoretically, for radiative processes and isospin-symmetry breaking.
The radiative corrections are routinely divided into the following: a nucleus-independent
part $\Delta^{\rm V}_{\rm R} = 2.361(38)$\%~\cite{[Mar06]}, a transition-dependent (Z-dependent) but
nuclear-structure-independent part $\delta_{\rm R}^\prime$~\cite{[Mar06],[Tow08]}
and a nuclear-structure-dependent part $\delta_{\rm NS}$~\cite{[Tow94],[Tow08]}.
The isospin-symmetry-breaking correction, $\delta_{\rm C}$, is defined through the following nuclear
matrix element:
\begin{equation}\label{fermime}
|\langle I=0, T\approx 1,
T_z = \pm 1 | \hat T_{\pm} | I=0, T\approx 1, T_z = 0 \rangle |^2
\equiv 2 (1-\delta_{\rm C}),
\end{equation}
where $\hat T_{\pm}$ are the raising and lowering bare isospin operators, respectively.

Application of the superallowed $\beta$-decay in testing the three-generations-quark
Standard Model of elementary particles requires both high-accuracy empirical
$ft$-values and a high-quality theory. The aim of this work is to present
new results on the $\delta_{\rm C}$ corrections obtained using
angular momentum and isospin projected density functional
theory (DFT) recently developed by our group~\cite{[Raf09c],[Dob09h],[Sat09sa],[Sat10s],
[Sat11sa],[Sat11sb],[Sat11sc]}.
The model will be introduced in Sec.~\ref{sec2}, the results
on  $\delta_{\rm C}$ will be presented in Sec.~\ref{sec3} and the paper
will be briefly summarized in Sec.~\ref{sec4}.

\section{The model}\label{sec2}

The degree of isospin mixing in atomic nuclei (isospin impurity) is, predominantly, a result of
subtle interplay between the short-range isospin-symmetry-conserving strong interaction and the long-range
isospin-symmetry-breaking Coulomb force. The Coulomb force polarizes wave functions of {\it all\/}
participating protons and in turn, neutrons, creating a number of conceptual and technical
difficulties within the nuclear shell-model (SM) having
rather profound consequences for calculations of $\delta_{\rm C}$
corrections~\cite{[Tow77],[Orm85y],[Har05b],[Tow08],[Tow10]}.
The SM can be used to compute only a part of the correction
related to configuration mixing $\delta_{\rm C1}$.
The second part, $\delta_{\rm C2}$, related to the radial mismatch of the single-particle
wave functions,  must be calculated independently using a
mean-field. Both corrections are treated as additive:
$\delta_{\rm C} = \delta_{\rm C1} + \delta_{\rm C2}$.
The consequences of this rather artificial division include, for example,
a necessity of using effective isospin operators which violate the SU(2) commutation rules.
This problem was recently noticed and discussed extensively by
Miller and Schwenk~\cite{[Mil09a]} who, however, didn't provide any quantitative estimate of the impact
of these deficiencies on the $\delta_{\rm C}$ results by
Towner and Hardy~\cite{[Tow08]} (TH) whose calculations set the standard in this field.

In contrast, Hartree-Fock (HF) and DFT
are free from these specific problems. Here all nucleons participate on equal footing and
a balance between long- and short-range effects is treated in a self-consistent manner.
It is well known, however, that these approaches break spontaneously
fundamental nuclear symmetries including rotational and isospin symmetry. In fact, the
isospin symmetry is violated both explicitly, by virtue of
charge-dependent interactions, and spontaneously which leads to unphysical
quenching of true isospin mixing.

In order to avoid spontaneous mixing and compute matrix element
(\ref{fermime}) in a fully quantum mechanical way using bare isospin operators we have recently developed
the angular-momentum and isospin
projected scheme on top of the Skyrme-DFT approach.
The approach is based on rediagonalization of the entire Hamiltonian including
the Coulomb interaction in good angular momentum and good isospin basis
\begin{equation}\label{ITbasis}
|\varphi ;\, IMK;\, TT_z\rangle =  
\hat P^T_{T_z T_z} \hat P^I_{MK} |\varphi \rangle ,
\end{equation}
projected of the self-consistent Slater determinant $|\varphi \rangle$.
Here $\hat P^T_{T_z T_z}$ and $\hat P^I_{MK}$ stand for the isospin
and angular momentum projection operators (see~\cite{[RS80]} for further
details).

For the Fermi matrix element (\ref{fermime})
the normalized bra and ket states are calculated in
the following way. The $| I=0, T\approx 1, T_z = \pm 1 \rangle$  state in
the even-even nucleus is projected of the self-consistent Slater determinant,
$|\psi \rangle$, representing the ground state in this nucleus:
\begin{equation}
  | I=0, T\approx 1, T_z = \pm 1 \rangle
  = \sum_{T\geq 1} c^{( \psi )}_{T} \hat P^T_{\pm 1, \pm 1}
     \hat P^{I=0}_{0,0} |\psi \rangle .
\end{equation}
The state $|\psi \rangle$ is unambiguously defined.
The  $|I=0, T\approx 1, T_z = 0 \rangle$, on the other hand,
is projected of the self-consistent Slater determinant, $|\varphi \rangle$,
representing the so-called antialigned configuration
$|\varphi \rangle \equiv |\bar \nu \otimes \pi \rangle$ (or  $| \nu \otimes
\bar \pi \rangle$), selected by placing the odd neutron and odd proton in
the lowest available time-reversed (or signature-reversed) single-particle
orbits:
\begin{equation}
  | I=0, T\approx 1, T_z = 0 \rangle
  = \sum_{T\geq 0} c^{( \varphi )}_{T} \hat P^T_{0 , 0}
     \hat P^{I=0}_{0,0} |\varphi \rangle .
\end{equation}
The selected single-particle configuration
$|\bar \nu \otimes \pi \rangle$ manifestly breaks the
isospin symmetry (see Fig.~1 in Ref.~\cite{[Sat10s]}). This is essentially the only way
to reach the $|T\approx 1, I=0\rangle$  states in odd-odd $N=Z$ nuclei. Indeed,
only the $T=0$ states in $N=Z$ nuclei can be represented by a single
Slater determinant. The final expression for the nuclear matrix element (\ref{fermime})
will be given in our forthcoming paper.

The two major drawbacks of the model in its present formulation include
({\it i\/}) the lack of pairing-correlations and ({\it ii\/}) the use of an old fashioned
and low-quality SV parametrization of the Skyrme force.
The latter deficiency pertains to the angular momentum projection which is
known to be ill-defined for density-dependent modern Skyrme and Gogny energy density
functionals (EDF)~\cite{[Ang01],[Rob07w],[Dob07sd],[Zdu07],[Lac09]}. At present,
the SV interaction augmented by a tensor term is essentially the only available
Skyrme interaction originating from the true Hamiltonian which can be safely used in connection with
the angular momentum projection without any further regularization~\cite{[Lac09]}.
It is worth mentioning here that the isospin impurities calculated using modern Skyrme forces in the isospin-projected
variant of our model~\cite{[Sat09sa],[Sat11sa]} are consistent with the recent data extracted from the Giant Dipole
Resonance decay studies in $^{80}$Zr~\cite{[Cor11]} and the isospin-forbidden E1 decay in
$^{64}$Ge~\cite{[Far03]}. This indicates that our model is in principle capable of
quantitatively capturing the amount of isospin mixing  that is important
in the context of making reliable calculations of $\delta_{\rm C}$ in spite of the fact that
$\delta_{\rm C}$ is mostly sensitive to the difference between the isospin impurities of parent and daughter
nuclei~\cite{[Har98],[Aue09]}.

\section{The results}\label{sec3}

The isospin-breaking corrections
$\delta_{\rm C}$ were computed by different groups and
with diverse nuclear models~\cite{[Dam69],[Tow08],[Sag96],[Lia09],[Aue09]}.
Our calculations were published recently in Ref.~\cite{[Sat11sc]} and compared
to the TH~\cite{[Tow08]}. Comparison of $\delta_{\rm C}$'s shows that
although individual values differ, both sets of calculations follow
a similar trend with increasing $A$. The differences between individual values
are stronger in light nuclei and are traced back to poor spectroscopic quality
of the SV parametrization. It is interesting to note that our results are
considerably larger as compared to the results of Ref.~\cite{[Lia09]} which are
based on the relativistic Hartree (RH) plus random phase approximation (RPA) formalism.

The average value of the nucleus-independent reduced life-time (\ref{ftnew})
calculated for twelve out of thirteen high-precision  superallowed $\beta$-decays
(excluding the $^{38}$K$\rightarrow ^{38}$Ar transition) equals
$\overline{{\cal F}t}$=3070.4(9)s, which is consistent with the CVC hypothesis.
This value was obtained using Gaussian-distribution-weighted formula to conform
with the standards of the Particle Data Group (PDG) and using our adopted $\delta_{\rm C}$
listed in Table~1 of Ref.~\cite{[Sat11sc]}. In the calculations we used the radiative corrections
and $ft$-values taken from Ref.~\cite{[Tow08]} and~\cite{[Tow10]},
respectively. The calculated $\overline{{\cal F}t}$ leads to $|V_{\rm ud}|=0.97447(23)$
which is in nice agreement with the TH result $|V_{\rm ud}^{\rm (TH)}|=0.97418(26)$ and
a central value deduced from the neutron decay $|V_{\rm ud}^{\rm (\nu )}|=0.9746(19)$.
Combining our $|V_{\rm ud}|$ with the results of $|V_{\rm us}|=0.2252(9)$
and $|V_{\rm ub}|=0.00389(44)$ adopted from the most recent PDG
compilation~\cite{[Nak10]} leads to $|V_{\rm ud}|^2 + |V_{\rm us}|^2
+ |V_{\rm ub}| $ = 1.00031(61). This result implies that the CKM unitarity inferred
from its first row is fulfilled with a 0.1\% precision.

The confidence level (CL) of our calculations can be assessed be performing the test
proposed recently in Ref.~\cite{[Tow10]}. The test indicates that the
CL of our model is lower than the CL of other models analyzed in Ref.~\cite{[Tow10]}.
It should be stressed, however, that the low CL of our model results primarily from
a very small $\delta_{\rm C}$ for a single transition in $A$=62.
This seems to be a part of deeper problem faced not only by our model but, in fact,
all models involving a self-consistent mean-field. Indeed, these models have problems in reproducing
quantitatively the rapid increase in $\delta_{\rm C}$ expected to occur already at $A\sim 62$,
see~\cite{[Tow10]}. In our model the increase in $\delta_{\rm C}$ takes place at
$A=70$, similar to the RH plus RPA calculations of Ref.~\cite{[Lia09]}. To lesser extent this problem is
also seen in the SM plus HF calculations of Ref.~\cite{[Har05b],[Orm95]}.

\begin{figure}
\begin{indented}
\item[]\includegraphics[width=0.7\columnwidth, clip]{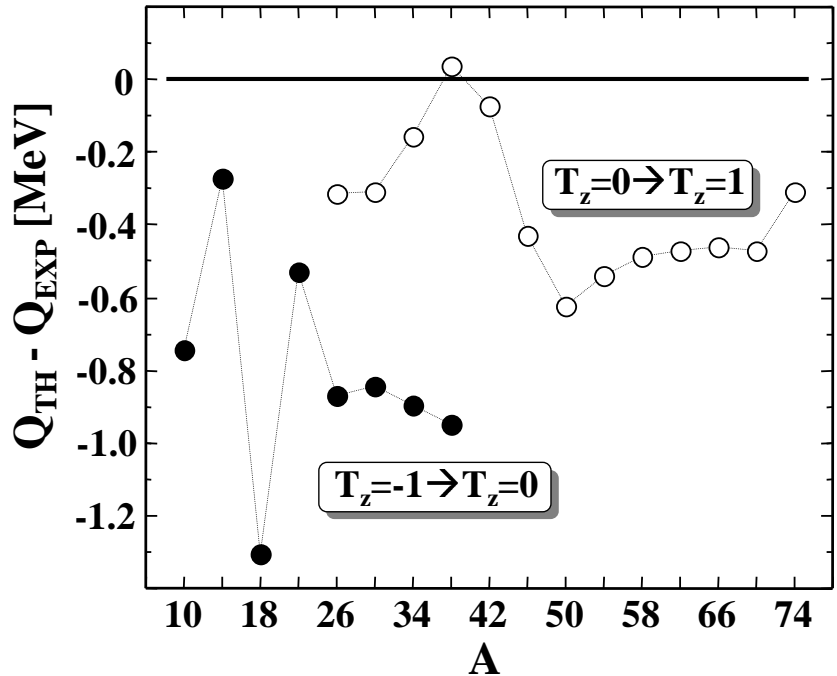}
\caption[T]{\label{fig-Qval}
Differences between theoretical and experimental $Q$-values for
the superallowed $\beta$-decays $I^\pi=0^+, T=1, T_z \rightarrow I^\pi=0^+, T=1, T_z + 1$ versus
the atomic number $A$. Black and white symbols represent transitions in mirror-symmetric pairs corresponding
to $T_z=-1$ and  $T_z= 0$, respectively.
}
\end{indented}
\end{figure}

The low confidence level of our results reflects poor spectroscopic quality
(ordering and energy of the single-particle levels) of the SV EDF.
Global characteristics of the functional are also unsatisfactory.
In particular, it reproduces the nuclear binding energies of even-even $T_z=\pm 1$ partners
(after projection) with accuracy of order of $1\% -2\%$ relative to the data.
As already mentioned $\delta_{\rm C}$ probes differences between the
partners and these quantities i.e. the relative binding
energies between the isobaric analogue states in the $\beta$-decay partners,
the $Q$-values, are captured quite well (see Fig.~\ref{fig-Qval}).
Nevertheless, even the $Q$-values show, in some cases, relatively large
deviations from the data and rather unacceptably
large fluctuations versus $A$, in particular, in light nuclei.
Part of these deviations reflects the well recognized deficiency of
conventional Skyrme-DFT models known in the literature under the name of the
Nolen-Schiffer anomaly, see~\cite{[Nol69],[Aue83],[Ben07x]}. In the calculations it manifests
itself as a discontinuity in the calculated relative $Q$-values
between the $T_z = -1 \rightarrow T_z = 0$ and  $T_z = 0 \rightarrow T_z = 1$ transitions
as shown in Fig.~\ref{fig-Qval}. Indeed, the figure shows that theoretical $Q$-values
tend to underestimate the experimental data and that the effect is larger
in the $T_z = -1 \rightarrow T_z = 0$ cases as compared to
the $T_z = 0 \rightarrow T_z = 1$ transitions.
At present it is neither clear how to cure this anomaly nor how to estimate
its influence on the isospin mixing.

\begin{figure}
\begin{indented}
\item[]\includegraphics[width=0.7\columnwidth, clip]{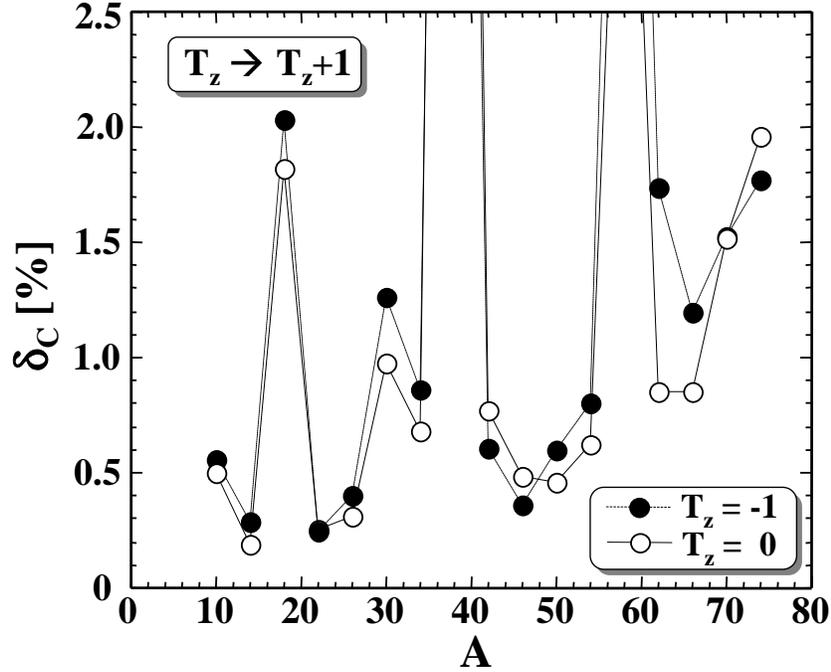}
\caption[T]{\label{fig-deltaC}
Isospin-mixing corrections $\delta_{\rm C}$ to the superallowed $\beta$-decays
$I^\pi=0^+, T=1, T_z \rightarrow I=0^+, T=1, T_z + 1$ versus the atomic number $A$.
Black and white symbols represent transitions in mirror-symmetric pairs corresponding
to $T_z=-1$ and  $T_z= 0$, respectively.
}
\end{indented}
\end{figure}

In order to check for the stability of our predictions we have carried out systematic calculations
of $\delta_{\rm C}$ for all $T_z = -1 \rightarrow T_z = 0$ and $T_z = 0 \rightarrow T_z = 1$
transitions from $A=10$ to $A=74$ being aware that most of these cases will probably
never be measured.
In the calculations we used $N=10$ and $N=12$ spherical harmonic oscillator shells
for $A\le 34$ and $A\ge 38$, respectively. Moreover, the anti-aligned states in $N=Z$ odd-odd nuclei
were calculated assuming signature-symmetry. The present results were calculated under different
conditions and are therefore different than our preferred values given in Ref.~\cite{[Sat11sc]}.
They are collected in Tab.~\ref{tab-deltaC} and visualized in Fig.~\ref{fig-deltaC}.

\begin{table}
\caption{\label{tab-deltaC} Calculated values of $\delta_{\rm C}$ for the superallowed $\beta$-decays
$I^\pi=0^+, T=1, T_z \rightarrow I^\pi=0^+, T=1, T_z + 1$ where $T_z = -1$ or $T_z = 0$.}
\begin{indented}
\item[]\begin{tabular}{@{}rrrcrrr}
\br
   &  \centre{2}{$\delta_{\rm C}$ [\%]} &  &  & \centre{2}{$\delta_{\rm C}$ [\%]} \\
   &  \crule{2}                         &  &  & \crule{2} \\
A  &  $T_z = -1$ & $T_z = 0$ & $\quad$ & A  &  $T_z = -1$ & $T_z = 0$ \\
\mr
10 &  0.559 & 0.497 & & 42 &  0.610 & 0.767 \\
14 &  0.290 & 0.189 & & 46 &  0.386 & 0.486 \\
18 &  2.031 & 1.819 & & 50 &  0.602 & 0.460 \\
22 &  0.243 & 0.255 & & 54 &  0.805 & 0.622 \\
26 &  0.399 & 0.308 & & 58 &  5.828 & 4.235 \\
30 &  1.260 & 0.974 & & 62 &  1.739 & 0.854 \\
34 &  0.865 & 0.679 & & 66 &  1.200 & 0.850 \\
38 &  8.315 & 9.826 & & 70 &  1.527 & 1.516 \\
   &        &       & & 74 &  1.768 & 1.956 \\
\br
\end{tabular}
\end{indented}
\end{table}

The calculations confirm earlier results concerning large mixing in
$A=18$ and pathologically large mixing in  $A=38$. Unusually large mixing
is also obtained in $A=58$. Note that these difficulties are seen in both
the $T_z = -1 \rightarrow T_z = 0$ and the $T_z = 0 \rightarrow T_z = 1$ cases.
In general, the values
of $\delta_{\rm C}$ in mirror-symmetric transitions
$T_z = -1 \rightarrow T_z = 0$ and $T_z = 0 \rightarrow T_z = 1$
follow a similar trend with increasing $A$ but the individual values
are different. In some cases the differences are sizeable.
It is interesting to observe that the largest differences between $\delta_{\rm C}$
in mirror-symmetric pairs occur in $A=62$ and $A=66$ i.e. exactly in the
region where the transition from small to large values of $\delta_{\rm C}$ is predicted
to occur in the SM plus Woods-Saxon potential calculations~\cite{[Tow08],[Tow10]}.
This result indicate that configuration mixing around $A=62$ is very fragile
in the self-consistent calculations. The effect may be sensitive to various
characteristics of the underlying EDF and/or to missing correlations
and requires further study.

\section{Summary and conclusions}\label{sec4}

We have extended the isospin and angular momentum projected DFT calculations
of the isospin breaking corrections $\delta_{\rm C}$ to all theoretically
possible superallowed $\beta$-decays in nuclei ranging from $A=10$ to $A=74$
in order to assess the stability of our predictions.
The calculations reveal that values of $\delta_{\rm C}$ in mirror-symmetric decays
$T_z = -1 \rightarrow T_z = 0$ and $T_z = 0 \rightarrow T_z = 1$ follow
a similar trend versus the atomic number $A$. The individual corrections
in mirror-symmetric decays are shown to be different, in some cases sizeably,
reflecting differences in configuration mixing in $T_z = \pm 1$ nuclei.
The largest differences were found for cases $A=62,66$ i.e.
in the region where the SM plus Woods-Saxon potential
calculations~\cite{[Tow08],[Tow10]} predict rapid
transition from small to large values of $\delta_{\rm C}$.
This transition is delayed in the self-consistent calculations.

\ack

This work was supported in part by the Polish Ministry of Science.
We acknowledge the CSC - IT Center for Science Ltd, Finland for the
allocation of computational resources.

\section*{References}


\end{document}